\documentclass[letterpaper,nofootinbib,showpac,onecolumn,eqsecnum,superscriptaddress,floatfix]{revtex4}

\usepackage{amsthm}
\usepackage{amssymb}   
\usepackage{dsfont}
\usepackage{mathtools} 
\usepackage{hyperref}
\hypersetup{colorlinks=true,linkcolor=blue,urlcolor=blue,citecolor=blue}
\usepackage{accents}
\usepackage{tensor}
\usepackage{xcolor}
\usepackage{graphicx}
\usepackage[cal=boondox]{mathalfa}
\usepackage{lipsum}
\usepackage{relsize}
\usepackage{color}
\usepackage{comment}
\usepackage{nameref}
\usepackage{lineno} 
\usepackage{orcidlink}
\usepackage{ulem}

\begin{document}


\title{Towards a de Sitter quintom-like cosmology from a reduction of a 7-dimensional $BF$ theory}
	
\author{Ricardo Escobedo\orcidlink{0000-0001-5815-4748}}
\email[Corresponding author:]{ricardo.escobedo@academicos.udg.mx}
\author{Roberto Santos-Silva\orcidlink{0000-0002-1265-0938}}
\email[]{roberto.santos@academicos.udg.mx}
\affiliation{Departamento de Ciencias Naturales y Exactas, CUValles,
Universidad de Guadalajara,
C.P. 46600, Ameca, Jalisco, M\'exico}

\author{Claudia Moreno\orcidlink{0000-0002-0496-032X}}
\email[]{claudia.moreno@academico.udg.mx}
\author{Rafael Hern\'andez-Jim\'enez\orcidlink{0000-0002-2740-9610}}
\email[]{rafael.hjimenez@academicos.udg.mx}

\affiliation{Departamento de F\'{i}sica, CUCEI, Universidad de Guadalajara, C.P. 44430, Guadalajara, Jalisco, M\'exico}

\date{\today}
	
\begin{abstract}
%

%
In this article, we analyze a 7-dimensional $BF$ theory that, upon dimensional reduction, transforms into an effective 4-dimensional action in which a generic metric is involved. A constraint is imposed in such a way that this is the Friedmann-Lema\^{i}tre-Robertson-Walker metric characterized by a scale factor $a(t)=\mathrm{exp}(H_0 t)$ with a Hubble constant $H_{0}=67.70 \,\rm [km\, s^{-1}\,Mpc^{-1}]$. We also compute the barotropic parameter $\omega \equiv p/\rho$ from our solutions. Three different scenarios are explored, in which the model transitions from quintessence to phantom regimes, crossing the theoretical phantom divide line. Consequently, the effective action represents a quintom-like scenario. Moreover, we derive some analytic solutions for the scale factor by treating it as a dynamical variable in the effective action. In this case, different outcomes are obtained, where the model can transit from a cosmological constant to a quintessence regime; however, it cannot cross the theoretical phantom divide line. However, this simplistic model successfully captures distinct phases of cosmic expansion.
\end{abstract}
	
\maketitle
	
\section{Introduction}



In this article, we explore a 7-dimensional model proposed in \cite{Mitskievich:2007pk}, which is based on a topological $BF$ theory with a Lagrangian independent of metric. It is well known that an $n$ dimensional $BF$ theory is a diffeomorphism invariant gauge theory and its action has the form
\begin{equation}
    S[A,B]=\int_{M}\mathrm{Tr}(B\wedge F[A]),
\end{equation}

where $B$ is a $(n-2)$-form valued in the Lie algebra of a group $G$, $A$ is a connection on a principal bundle $P$ over an arbitrary $n$-dimensional manifold $M$, $F:=dA+A\wedge A$ is the curvature of this connection, and the trace is taken with respect to the non-degerate Killing-Cartan metric of the given group $G$. The particular form of the action is the main reason why these models are called $BF$ theories~\cite{Celada_2016}. These theories have been studied in \cite{PhysRevD.78.064046}, and define a topological theory; namely, with no degrees of freedom\cite{cmp/1104179527,BIRMINGHAM1991129,CAICEDO1995292,Cattaneo_Cotta}.

One of the key concepts in this work is to split the 7-dimensional space, and through a dimensional Kaluza-Klein \cite{Efremov:2010zz} reduction on homological $3$-dimensional spheres, derive an effective $4$-dimensional theory. 
%
%
%
%
The introduction of the $BF$ theory was postulated in 1977 by Pleba\'{n}ski~\cite{10.1063/1.523215}. It is well known that General Relativity (GR) in four dimensions can be formulated as a modified $BF$ action, achieved by adding a constraint (known as the simplicity constraint) to the fields $B$. Solving this condition leads to the auto-dual formulation of GR, ensuring that the theory acquires degrees of freedom (see also~\cite{Capovilla_2001,Celada_2016,R_De_Pietri_1999}). Thus, the field $B$ can be considered as a fundamental variable instead of the metric $g_{\mu\nu}$~\cite{PhysRevLett.63.2325}. Moreover, in higher dimensions $n\geq 4$, and in the absence of a cosmological constant $\Lambda$, GR can also be formulated as a constrained $BF$ theory~\cite{Freidel,Merced_Escobedo}. 

One of the main applications of the relativistic theory of gravity is cosmology, as the most promising framework for making predictions that could confirm the theory. Observational cosmology is advancing rapidly, and it is expected that this can be evidence for testing a quantum gravity theory (for instance loop quantum gravity), particularly through cosmic background radiation~\cite{RovBook}. 

The multi-field cosmology paradigm has been proven to be an effective framework for incorporating various significant features and components of the universe within a unified model. Several approaches, such as the formalism for quantum fluctuations \cite{Sasaki:1995aw}, the scalar field dynamics and nonlinear effects \cite{Liddle:1998jc,Rigopoulos:2002mc}, reviews on inflationary models \cite{Bassett:2005xm,Wands:2007bd}, 
multi-field inflation and its impact on the cosmic microwave background \cite{Lalak:2007vi}, quantum corrections \cite{Ashoorioon:2009wa}, and the effects of heavy fields on inflationary trajectories \cite{Achucarro:2010jv,Achucarro:2010da}, are related with the phenomenon of early acceleration, known as inflation. Further developments in this topic are the non-canonical kinetic terms and sharp features in the potential \cite{Achucarro:2012sm,Achucarro:2012yr}, quantum entanglement in multi-field inflation \cite{Pi:2012gf}, non-Gaussianities in curved field space models \cite{Renaux-Petel:2015mga}, swampland constraints and supergravity \cite{Brown:2017osf,Achucarro:2017ing}, heavy field interactions with inflationary fluctuations \cite{Achucarro:2018vey}, reheating mechanisms and energy transfer from the inflaton to Standard Model particles \cite{Aragam:2020uqi}, and the role of dark matter \cite{Hu:2000ke, Hui:2016ltb, Tellez-Tovar:2021mge}. Additionally, the phenomenon of late cosmic acceleration has been widely studied within various theoretical frameworks. These include modifications to General Relativity as a potential explanation \cite{Boyle:2001du}, scalar field models driving acceleration \cite{Kim:2005ne,vandeBruck:2009gp}, the role of modified gravity theories in reproducing late-time expansion \cite{BeltranJimenez:2012xud,Vardanyan:2015oha}, the dynamics of interacting dark energy models \cite{Leithes:2016xyh}, the interplay between inflationary initial conditions and late-time acceleration \cite{Akrami:2017cir}, string theory-inspired scenarios highlighting their implications for cosmic acceleration \cite{Cicoli:2020cfj,Cicoli:2020noz}, constraints on dark energy evolution using observational data \cite{Akrami:2020zfz}, cosmological solutions in modified theories of gravity \cite{Socorro:2020nsm}, dynamical systems approaches to dark energy \cite{Paliathanasis:2021fxi}, swampland constraints on late acceleration \cite{Burgess:2021qti,Burgess:2021obw}, and the role of extra-dimensional effects in shaping cosmic expansion \cite{Anguelova:2021jxu}.

Regarding early and late acceleration, achieving the crossing of the phantom divide line is a highly sought-after characteristic in scalar-field cosmology. It has been shown that accomplishing this crossing was not possible when focused solely on a single scalar field or fluid (unless stability is not a requirement)~\cite{Cai:2009zp}. The phantom divide line is a theoretical boundary that separates two different scenarios of cosmic acceleration based on the equation of state parameter of dark energy: defined by the ratio of the pressure-to-energy density $\omega \equiv p/\rho$. On the one hand, the quintessence regime occurs when $\omega > -1$ but still less than 0, corresponding to a scenario where cosmic expansion is accelerating, but at a decreasing rate. On the other hand, the phantom framework occurs when $\omega<-1$, here the expansion of the universe is accelerating at an increasing rate. This phantom stage can be explained by the presence of exotic forms of dark energy with repulsive properties that defy conventional physics. Moreover, the standard quintom scenario~\cite{Feng:2004ad} considers two scalar fields, a quintessence and a phantom, to achieve this crossing in a straightforward manner. As a result, quintom fields allow, in specific instances, the circumvention of the initial singularity through a ``big-bounce''~\cite{Socorro:2022aoz} (see also~\cite{Cai:2009zp}). Notice also that generalized quintom cosmology has been proposed in~\cite{PhysRevD.70.043539}.

The outline of the paper is as follows. In Section~\ref{Effective action} we review the 7-dimensional theory and set the definitions and conventions. Section~\ref{Equations of Motion} is devoted to the effective four-dimensional Lagrangian density using dimensional reduction (Kaluza-Klein). In section~\ref{Example} we present a toy model described by a homogeneous and isotropic flat Friedmann-Lema\^{i}tre-Robertson-Walker (FLRW) metric, here we compute and study the equation of motion (EOM) of the theory. In Section~\ref{Energy-momentum tensor} we obtain the energy-momentum tensor of a perfect fluid to analyze our cosmological model via the barotropic parameter $\omega$. In Section~\ref{scale_factor} we consider the case where the scale factor is a dynamical variable, then we compute a solution for such scale factor and the fields involved using series. Finally in Section~\ref{Conclusions} we present our concluding remarks. Moreover, to be self-contained in Appendix \ref{Appendix_A} we write the definitions of differential forms used throughout the paper. In Appendix \ref{Appendix_B}  we show an analytic solution for the scale factor and the fields involved for a weak-field approximation, and Appendix \ref{pressure_density} is a summary of the computations used to compute the barotropic factor.

\section{Effective action}
\label{Effective action}

In this section, we present a brief review following references \cite{Mitskievich:2007pk,Efremov:2010zz}. Also, we have included an Appendix~\ref{Appendix_A} on differential forms, in which we provide essential definitions that will be used throughout this paper.
%
%

The $7$-dimensional $BF$ theory is given in the following way~{\footnote{In this work we are working with the reduced Planck units, i.e.,  $8\pi G=1$ and $c=1$.}}
\begin{equation}
    S_7=m_0 k \int_{X_3 \times M_4} B \wedge F \,,
    \label{action}
\end{equation}
where $m_0$ is the mass scale parameter, $k$ is the coupling constant of the theory and the 7-dimensional manifold is regarded as $X_3 \times M_4$, here $X_3$ is a $3$-dimensional non-trivial topology graph manifold which consists only of Seifert fibered spheres $M^I$, $I=1,...,m$, glued together with $2$-torus (this process is named plumbing), $B$ and $F$ are $3$-form and $4$-form respectively defined over the whole $7$ dimensional manifold. 
Each of these spheres is classified by the following Seifert invariants $\{ (a^I_1,b^I_1) ,\ldots,(a^I_{\ell_I},b^I_{\ell_I}) \}$ \cite{Mitskievich:2007pk}. Moreover, the Seifert invariants are classified by the Euler rational number (we consider only tree-type graph manifolds plumbing Seifert spheres),
\begin{equation}
    e^I=- \sum_{i=1}^{\ell_I} \frac{b^I_i}{a^I_i} \,.
\end{equation}
The reduced plumbing matrix is defined as follows:
\begin{eqnarray}
    K^{II}= e^I \,, \nonumber \\
    K^{IJ}= \frac{1}{p^{IJ}} \,,
\end{eqnarray}
where $p^{IJ}$ is the fiber intersection number in the torus that connects the spheres $M^{I}$ and $M^J$. The Kaluza-Klein dimensional reduction \cite{efremov2014universe} was carried out in the action \eqref{action} by assuming that the $B$ and $F$ forms are expressed as:

%
\begin{eqnarray}\label{ansatz}
     B =\left(B_I \otimes \kappa^I - \frac{1}{m_0 k} F_D^I \otimes \sigma_I \right)+\left(F_I \otimes \kappa^I + \frac{1}{m_0 k} dH_D^I \otimes \sigma_I \right) \,, \nonumber \\
     F =d\left(A_I \otimes d \kappa^I + \frac{1}{m_0 k} H_D^I \otimes d \sigma_I \right) \,, 
\end{eqnarray}
where $d$ is the exterior derivative, $F_I=d A_I$, $F_D^I$, $B_I$ are $2$-forms and $H_D^I$ is a $1$-form over $M_4$. Finally, $\kappa^I$ and $\sigma_I$ are the $U(1)^m$ connection $1$-forms and their, respectively, dual $1$-forms, such that they satisfy:
\begin{equation}\label{condition}
    \int_{X_3} \sigma_I \wedge d \kappa^J=\int_{X_3} d\sigma_I \wedge \kappa^J = \delta_I^J \,,
\end{equation}
where $\delta^{J}_{I}$ is the 4-dimensional Kronecker delta. Here, the set of forms $\kappa^J$ and $\sigma_I$ followed by the condition Eq.~\eqref{condition} is known as a symplectic base. Now, let us consider the following choice due to the gauge fixing~\cite{efremov2014universe}
\begin{equation}
    F_D^I =\frac{1}{2} K^{IJ} \star F_J, \qquad \qquad H_D^I =\frac{1}{2} K^{IJ} \star H_J \,,
\end{equation}
where $\star$ is the Hodge star operator whose explicit definition can be checked in Appendix~\ref{Appendix_A}. Assuming that $M_4$ has no boundary; and without loss of generality we consider only one element $K^{11}=1$, thus after integrating the forms on $X_3$ (graph manifold), the 7-dimensional $BF$ action Eq.~\eqref{action} on the $4$-dimensional space-time $M_4$ is reduced to (see details in~\cite{efremov2014universe}):
\begin{equation} \label{faction}
    S[A, B, g]=\int_{M_4}\left[ m_0 k A \wedge dB-\frac{1}{2}dA\wedge \star dA-\frac{1}{2}dB\wedge \star dB-\frac{1}{4m_0 k}\star dA \wedge d \star dB \right] \,,
\end{equation}
where $A$ is a $1$-form and $B$ is a $2$-form defined over the space-time. It is worth to stress that the presence of the Hodge dual operator in the effective action implies that it depends explicitly of the metric. Notice also that $A$ is not a connection variable and that there is no curvature tensor of the $2$-form involved here~\cite{efremov2014universe}.

\section{Equations of Motion (EOM)}
\label{Equations of Motion}

This section expresses the effective action from Eq.~\eqref{faction} using the inner product, wedge product, and Hodge star operator $\star$ (refer to Appendix~\ref{Appendix_A}), to simplify the derivation of the EOM. Recalling that the inner product in the space of $p$-forms is defined by (see Appendix \ref{Appendix_A} for explicit definitions):
\begin{equation}
(\alpha,\beta):=\int_{M_4}\alpha \wedge \star \beta \,,
\end{equation}\label{effective_action}
we can rewrite each term of the effective action Eq.~\eqref{faction} as follows:
\begin{eqnarray}
\label{1}\int_{M_4} m_0 k A\wedge dB&=&(-1)^4 m_0 k \int A\wedge \star^2 dB=m_0 k (A, \star dB) \,, \\
\label{2}-\frac{1}{2}\int_{M_4}dA \wedge \star dA &=& -\frac{1}{2}(dA,dA) \,, \\
\label{3}-\frac{1}{2}\int_{M_4}dB \wedge \star dB&=&-\frac{1}{2}(dB,dB) \,, \\
\label{4}-\frac{1}{4m_0 k}\int_{M_4}\star dA \wedge d\star B &=& -\frac{1}{4m_0 k}\int_{M_4}d\star dB \wedge \star dA =-\frac{1}{4m_0 k}(d\star dB, dA) \,.
\end{eqnarray}
Using these expressions, the effective action is redefined as:
\begin{equation}
S[A, B, g]= m_0 k (A, \star dB)-\frac{1}{2}(dA, dA)-\frac{1}{2}(dB,dB)-\frac{1}{4m_0 k}(d\star dB, dA) \,.
\end{equation}
It should be noted that to compute the EOM related to the $1$-form $A$, it is convenient to rewrite the action using the properties of the inner product $(\cdot,\cdot)$ (see Appendix \ref{Appendix_A}), resulting in the following equation:
\begin{equation}\label{action_rewritten}
    S[A, B, g]=m_0 k (A, \star dB)-\frac{1}{2}(A, d^{\dagger} dA)-\frac{1}{2}(B, d^{\dagger}dB)-\frac{1}{4m_0 k}(A, d^{\dagger}d \star dB) \,,
\end{equation}
where $d^\dagger$ is the adjoint operator of the differential $d$. 
Now, we can compute the EOM by performing a variation with respect to the fields $A$ and $B$. The variation of the previous action (Eq.~\eqref{action_rewritten}) with respect to the field $A$ reads as follows:
\begin{equation}
    \delta_A S=(\delta_A A, m_0 k \star dB-d^{\dagger}dA-\frac{1}{4m_0 k}d^{\dagger}d \star dB)=0 \,,
\end{equation}
resulting in this equation:
\begin{equation}\label{EOM_1}
m_0 k \star dB-d^{\dagger}dA-\frac{1}{4m_0 k}d^{\dagger}d \star dB=0 \,.
\end{equation}
We take the previous procedure to compute the variation with respect to the $2$-forms $B$, but before it is convenient to rewrite Eqs.~\eqref{1} and \eqref{4} in terms of 
derivatives, that is,
\begin{eqnarray}
m_0 k \int_{M_4}A \wedge dB &=& -m_0 k (B, \star dA) \,, \\
(d\star dB, dA)&=& -(B, d^{\dagger}d\star dA) \,.
\end{eqnarray}
The effective action is then revised to:
\begin{equation}
S[A, B, g] = -m_0 k (B, \star dA)-\frac{1}{2}(dA, dA)-\frac{1}{2}(B, d^{\dagger}dB)+\frac{1}{4m_0 k}(B, d^{\dagger}d\star dA) \,.
\end{equation}
Now, the variation with respect to the field $B$ yields
\begin{equation}
\delta_B S= (\delta B, -m_0 k \star dA-d^{\dagger}dB+\frac{1}{4m_0 k}d^{\dagger}d\star dA)=0 \,,
\end{equation}
resulting in the following equation:
\begin{equation}\label{EOM_2}
-m_0 k \star dA-d^{\dagger}dB+\frac{1}{4m_0 k}d^{\dagger}d\star dA=0 \,.
\end{equation}
Our final EOM are given by Eqs. $\eqref{EOM_1}$ and $\eqref{EOM_2}$. We can observe that this is a coupled system of equations that needs a bit of algebraic manipulation. In the next subsection, we will show how to massage it. 

\subsection{Analysis of EOM}
\label{Manipulation of EOM}

In this subsection, we manipulate our EOM to decouple the system of equations obtained in the previous section. First, we work with Eq.~\eqref{EOM_2}, which is rewritten as:
\begin{equation}\label{EOM_2'}
    d^{\dagger}dB= \left( \frac{1}{4m_0 k}d^{\dagger}d-m_0 k \right)\star dA \,.
\end{equation}
On the other hand, we apply $\star d$ in Eq.~\eqref{EOM_1} obtaining the following expression:
\begin{equation}
    m_0 k \star d \star dB-\star d \star d (\star dA)-\frac{1}{4m_0 k}\star d \star d \star d \star dB=0 \,.
    \label{newex}
\end{equation}
Now, we plug in Eq.~\eqref{EOM_2'} into Eq.~\eqref{newex}, yielding:
\begin{eqnarray}
    m_0 k \left( \frac{1}{4m_0 k}d^{\dagger}d-m_0 k \right)\star dA-d^{\dagger}d(\star dA)-\frac{1}{4m_0 k}d^{\dagger}d\left( \frac{1}{4m_0 k}d^{\dagger}d-m_0 k \right)\star dA=0 \,.
\end{eqnarray}
Note that the equation depends only on the $1$-form $A$. We simplify the above formula to get the expression:
\begin{equation}
16m_0^4 k^4 (\star dA)+8m_0^2 k^2 d^{\dagger}d (\star dA)+d^{\dagger}dd^{\dagger}d(\star dA)=0 \,,
\label{simplify}
\end{equation}
which is the square of the Klein-Gordon equation for the field ($1$-form) $\star d A$, then Eq.~\eqref{simplify} becomes
\begin{equation} \label{kgA}
\left( d^{\dagger}d+4m_0^2 k^2 \right)^2 (\star d A)=0 \,.
\end{equation}
Similarly, we obtain the EOM of the 2-form $B$:
\begin{equation}
\label{kgb}
    \left( d^{\dagger}d+4m_0^2 k^2 \right)^2 (\star d B)=0 \,.
\end{equation}
Note that we can first solve (without loss of generality) the equation for $\star d  A$ and substitute it into the EOM to find $B$. Thus, we introduce a master equation to handle the system more easily, that is,
\begin{equation}\label{quadratic_equation}
\left( d^{\dagger}d+4m_0^2 k^2 \right)^2 E=0 \,,
\end{equation} 
here, we define:
\begin{equation} \label{E_def}
E :=\star d A \;\text{or} \, \star dB  \,.\end{equation}
This expression leads to finding four solutions, since it is the square of a second-order equation. Once we find $E$, we integrate it to get $A$, we then substitute it into Eq.~\eqref{EOM_2} to obtain the solution of the field $B$.

\section{Toy model}
\label{Example}

In order to study a simplified cosmological model, we choose the particular de Sitter universe described by the Friedmann-Lema\^{i}tre-Robertson-Walker (FLRW) metric given by:
\begin{equation}\label{RW1}
ds^{2}= dt^{2}-a(t)^{2}\delta_{ij}dx^{i}dx^{j}\,,
\end{equation}
where the scale factor  is $a(t)=\mathrm{exp}(H_0 t)$, $t$ stands for the cosmological time, and $H_0$ is the Hubble constant\footnote{Latin indices run from $1$ to $3$.}. To do so, we impose that the metric of the effective action Eq.\eqref{faction} be constrained by the FLRW metric with the conditions mentioned above. Thus, the departure point of our analysis is given by the following action:
\begin{eqnarray}\label{action_constrained}
    & &S[A, B, g, \lambda, \mu,\nu]=\int_{M_4}\left[ m_0 k A \wedge dB-\frac{1}{2}dA\wedge \star dA-\frac{1}{2}dB\wedge \star dB-\frac{1}{4m_0 k}\star dA \wedge d \star dB \right.\nonumber\\
    &&  -\lambda (g_{00}-1)-\mu^{i}g_{0i}-\nu^{ij}(g_{ij}+a(t)^{2}\delta_{ij}) \bigg]  \,.
\end{eqnarray}
where $\lambda$, $\mu^{i}$, $\nu^{ij}$ are Lagrange multipliers, $a(t)=\mathrm{exp}(H_0 t)$ is the particular scale factor imposed and $\delta_{ij}$ is the 3-dimensional Kronecker delta. Recall that the metric is in the Hodge dual operator which is given explicitly in Appendix \ref{Appendix_A} in Eq.\eqref{hodge}. Subsequently, solving the equations of motion of the Lagrange multipliers $\lambda$, $\mu^{i}$ and $\nu^{ij}$ then, substituting back into the action Eq.\eqref{action_constrained} and also demanding that the fields $A$ and $B$ are only functions of time $t$ since the universe that we want to study is isotropic and homogeneous, lead to:
\begin{eqnarray}
    & &S[A,B] = \int_{M_4} \left[ -A_1 \left(\partial_0 B_{23}\right)+ A_2 (\partial_0 B_{13}) - A_3 (\partial_0 B_{12}) + \frac{a}{4}(\partial_0 A_1)^2+ \frac{a}{4}(\partial_0 A_2)^2+\frac{a}{4}(\partial_0 A_3)^2  - \frac{1}{2a}\left(\partial_0 B_{12}\right)^2  \right. \nonumber \\  & & - \frac{1}{2a}(\partial_0 B_{13})^2 -  \frac{1}{2a}(\partial_0 B_{23})^2 - \frac{1}{4 m_0 k a}(\partial_0 A_3) \partial_0\left(\frac{1}{a}\partial_0 B_{12}\right) + \frac{1}{4 m_0 k a}\left(\partial_0 A_2\right) \partial_0\left(\frac{1}{a}\partial_0 B_{13}\right) \nonumber \\ & & \left.- \frac{1}{4 m_0 k a}(\partial_0 A_1) \partial_0\left(\frac{1}{a}\partial_0 B_{23}\right) \right]  \,.
\end{eqnarray}

On the other hand, to solve Eq. \eqref{quadratic_equation}, we first need to write it explicitly in coordinates; being this equation the square of the Klein-Gordon equation, then, first we solve the equation \footnote{Notice that this is not the only solution since the EOM can be regarded as a nilpotent operator $K=\left( d^{\dagger}d+4m_0^2 k^2 \right)$, that satisfies $K^2=0$. A set of solutions is $K E=0$.}
\begin{equation}\label{Klein_Gordon_eq}
    \left( d^{\dagger}d+4m_0^2 k^2 \right)E=0 \,.
\end{equation}
Then, using $d^\dagger = (-1)^{n(p+1)}\star d \star$, the definition of the Hodge star operator Eq. \eqref{hodge}, considering that the fields are only functions of cosmological time $t$, and bearing in mind that the metric satisfies the constraint given above, the Klein-Gordon Eq.~\eqref{Klein_Gordon_eq} for $E$ being a two-form (this is for case $E=\star d A$), collapses to
\begin{eqnarray} 
&&\left[a \partial_0 \left(\frac{1}{a} \partial_0 E_{12} \right)+4 m_0 k^2 E_{12} \right] dx^1 \wedge dx^2 + \left[a \partial_0 \left(\frac{1}{a} \partial_0 E_{13} \right)+4 m_0 k^2 E_{13} \right] dx^1 \wedge dx^3   \\
&&+ \left[a \partial_0 \left(\frac{1}{a} \partial_0 E_{23} \right)+4 m_0 k^2 E_{23} \right] dx^2 \wedge dx^3 + 4 m_0 k^2 (E_{01} dx^0 \wedge dx^1+E_{02} dx^0 \wedge dx^2+E_{03} dx^0 \wedge dx^3)=0 \,. \nonumber
\end{eqnarray}
Each term in the brackets is linearly independent, implying that the components $E_{01}, E_{02}$ and $E_{03}$ must vanish due to the linearity of the $2$-form basis, which in turn indicates that $A_0=0$. Then, from Eq.~\eqref{Klein_Gordon_eq} we obtain three differential equations for the spatial directions; however, note that these equations are the same for every component of the field due to the homogeneity and isotropy observed in the space-time model, yielding
\begin{eqnarray}\label{geom}
    a \partial_0 \left(\frac{1}{a} \partial_0 E_{12} \right)+4 m_0 k^2 E_{12}=0 \,,  \\
    a \partial_0 \left(\frac{1}{a} \partial_0 E_{13} \right)+4 m_0 k^2 E_{13}=0 \,,  \\
    a \partial_0 \left(\frac{1}{a} \partial_0 E_{23} \right)+4 m_0 k^2 E_{23}=0 \,.
\end{eqnarray}
Therefore, we only need to solve one of these equations since the other directions are the same, thus without loss of generality we determine the component $E_{12}$, which has:
\begin{equation}
   \ddot{E}_{12} - \frac{\dot{a}}{a} \dot{E}_{12}+ 4 m_0^2 k^2 E_{12}=0 \,.
\end{equation}
Now, to solve the above equation and compare it with a cosmological model, we substitute the scale factor $a(t)=\mathrm{exp}(H_0 t)$ into the previous formulas and solving for $E_{12}$ we obtain the following expressions:
\begin{eqnarray}
\label{Sol_1} E_{12} =  e^{\frac{1}{2} (H_0 - \sqrt{H_0^2 - 16 k^2 m_0^2}) t} \,,  \\ 
\label{Sol_2} E_{12}=e^{\frac{1}{2} (H_0 + \sqrt{H_0^2 - 16 k^2 m_0^2}) t } \,.
\end{eqnarray}
Recall that we are dealing with a $4$-th order system (since the master EOM is the square of the Klein-Gordon equation), we have another two solutions given by
\begin{eqnarray}
    \label{Sol_3} E_{12}=-\frac{e^{\frac{1}{2} (H_0 -\sqrt{H_0^2 - 16 k^2 m_0^2}) t} (1 + \sqrt{H_0^2 - 16 k^2 m_0^2} \,t)}{H_0^2 - 16 k^2 m_0^2} \,, \\
    \label{Sol_4 }E_{12}=\frac{e^{\frac{1}{2} (H_0 +\sqrt{H_0^2 - 16 k^2 m_0^2}) t} (-1 + \sqrt{H_0^2 - 16 k^2 m_0^2} \,t)}{H_0^2 - 16 k^2 m_0^2} \,. \label{solution}
\end{eqnarray}
Thus, we can easily compute the component $A_3$ from Eq.~\eqref{E_def} as follows:
\begin{equation}
    A_3 = -\int \frac{1}{a(t)} E_{12}(t) dt \,.
\end{equation}
Moreover, a second formula is obtained from Eq.~\eqref{solution}, having $A_3$ as:
\begin{equation}\label{A_3}
  A_3 = \frac{2 e^{\frac{1}{2} (-H_0 + \sqrt{H_0^2 - 16 k^2 m_0^2}) t }\left[3 \sqrt{H_0^2 - 16 k^2 m_0^2} 
    +(-H_0^2  + 16 k^2 m_0^2) t + H_0 (-1 + \sqrt{H_0^2 - 16 k^2 m_0^2} t)\right]}{(H_0^2 - 
   16 k^2 m_0^2) \left(H_0 - \sqrt{H_0^2 - 16 k^2 m_0^2}\right)^2} \,.
\end{equation}
Then, from Eq.~\eqref{EOM_2}, $B$ can be integrated directly in terms of $\star dA$, hence explicitly the component $B_{12}$ is:
\begin{equation}
    B_{12} = -2 k m_0 \int \left( a(t) \int \left(\frac{1}{a(t)} E_{12}(t)  dt \right) dt \right) 
    + \frac{1}{4 k m_0} E_{12}(t) \,,
\end{equation}
therefore, we have:
\begin{eqnarray}\label{B_12}
   B_{12}  =\frac{e^{\frac{1}{2} (-H_0 + \sqrt{H_0^2 - 16 k^2 m_0^2}) t} \left[H_0^2 + 
   2 k^2 m_0^2 (-11 + 3 \sqrt{H_0^2 - 16 k^2 m_0^2} t)\right]}{8 k^3 m_0^3 (H_0^2 - 
   16 k^2 m_0^2)} \,.
\end{eqnarray}
Finally, since the EOM are expressed in terms of differential forms, the relationships among the other components can be established as follows:
\begin{eqnarray}
    \label{C1} A_1=A_3, \quad A_2=A_3 \,, \\
    \label{C2} B_{13}=-B_{12}, \quad B_{23}=-B_{12} \,, \\
    \label{C3} B_{01}=B_{02}=B_{03}=0 \,.
\end{eqnarray}
Once we have the full set of solutions, we compute the energy-momentum tensor of a perfect fluid in the next section. 

%
\section{Energy-momentum tensor}
\label{Energy-momentum tensor}
%

\begin{figure}[h]
\centering
\includegraphics[width=0.5\textwidth]{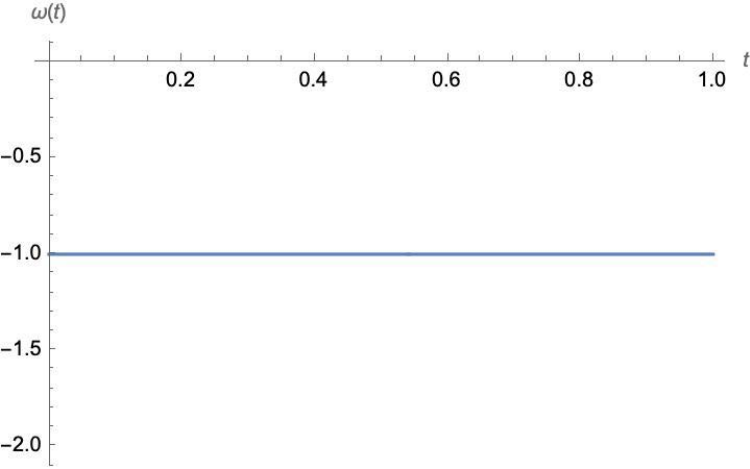}
\caption{Time evolution of the barotropic parameter $\omega (t)$ with mass scale parameter $m_{0}=0.727$. Here we consider solution $y_{1}$ (Eq.~(\ref{solutionA}). Note that $\omega (t)\simeq -1$ which corresponds to a cosmological constant setting. Note that time is measured in reduced Planck units since $8\pi G=1$.}
\label{fig:w_m_0}
\end{figure}
\begin{figure}[h]
\includegraphics[width=0.49\textwidth]{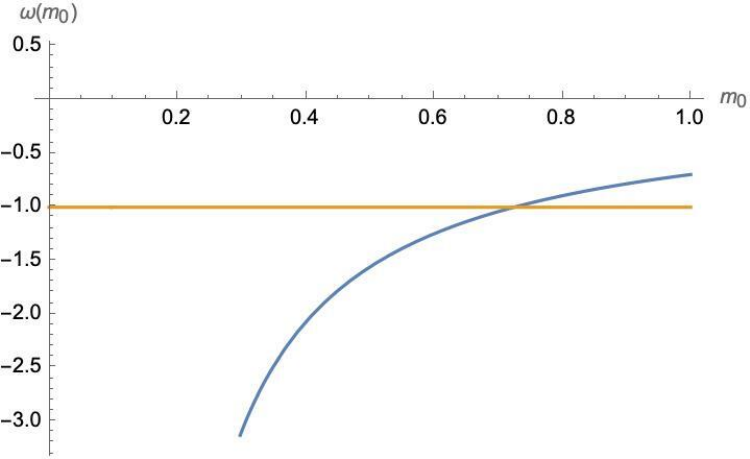}
\caption{In this figure we show the barotropic parameter $\omega (m_{0})$ in terms of the mass $m_{0}$ at $t=0$. Here we consider solution $y_{1}$ (Eq.~(\ref{solutionA}). Note that $\omega (m_{0})$ can cross the phantom line (orange solid line) and for larger $m_{0}$ it tends to the quintessence scenario.}
\label{fig:w_m_1}
\end{figure}
\begin{figure}[h]
\includegraphics[width=0.5\textwidth]{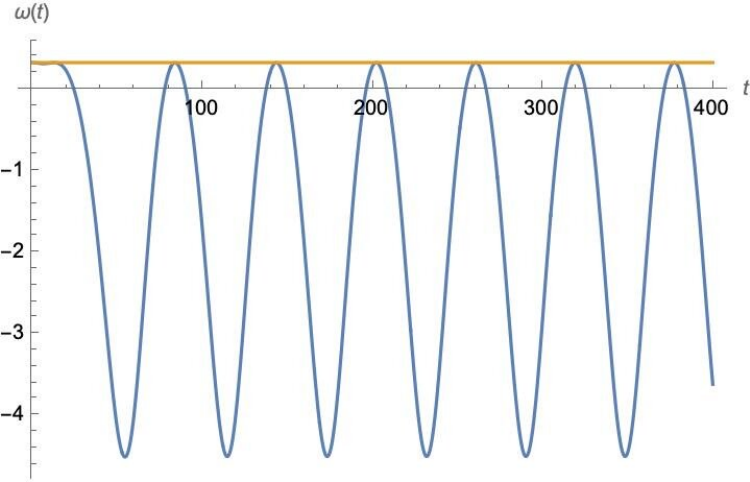}
\centering
\caption{Time evolution of the barotropic parameter $\omega (t)$ with mass scale parameter $m_{0}=0.027$. Here we consider solution $y_{2}$ (Eq.~(\ref{solutionB})). Note that $\omega (t)$ oscillates from positive to negative values, it does transit from the quintessence to the phantom regimes; moreover, its maximum value reaches $\omega \simeq 1/3$ that corresponds to a radiation fluid. Note that time is measured in reduced Planck units since $8\pi G=1$.} 
\label{fig:w_m_10}
\end{figure}

We consider the effective action Eq.~\eqref{faction} as the matter component. Hence, we can calculate its corresponding energy-momentum tensor $T^{\mu\nu}$ by the variation of the metric $g_{\mu \nu}$\footnote{Greek index run from 0 to 3}:
\begin{equation}\label{definition_EMT}
    \frac{\delta S}{\delta g_{\mu\nu}}=-\frac{1}{2}(-g)^{1/2}T^{\mu\nu}d^{4}x \,,
\end{equation}
where $g$ is the determinant of the metric. We point out here that in the effective action Eq.\eqref{faction} we will impose that the fields $A$ and $B$ have the form given previously in Eq.\eqref{A_3} and Eq. \eqref{B_12} with the conditions Eq.\eqref{C1}-Eq.\eqref{C3}. Therefore, the action in Eq.\eqref{faction} depends only on the metric $g_{\mu\nu}$. Bearing in mind these conditions, it turns out that the variation of the action Eq.\eqref{faction} 
with respect to the metric reads as:
\begin{eqnarray}\label{variation_metric}
    \frac{\delta S[g]}{\delta g_{\mu\nu}}&=&-\frac{1}{4}g^{\mu\nu}\bigg[ dA\wedge \star dA+dB\wedge \star dB+\frac{1}{2m_0 k}\star dA\wedge d\star dB \bigg]-\frac{1}{8m_0 k}\star dA\wedge d(g^{\mu\nu}\star dB) \nonumber\\
    &+&\frac{1}{4}(-g)^{1/2}\partial_{[\alpha}A_{\beta]}g^{\alpha(\mu}g^{\nu)\lambda}g^{\beta\rho}\epsilon_{\lambda \rho \gamma \delta}\bigg( dA\wedge dx^{\gamma}\wedge dx^{\delta}+\frac{1}{2m_0 k}dx^{\gamma}\wedge dx^{\delta}\wedge d\star dB \bigg) \nonumber\\
    &+&\frac{1}{12}(-g)^{1/2}\partial_{[\alpha}B_{\beta\gamma]}g^{\alpha(\mu}g^{\nu)\delta}g^{\beta\lambda}g^{\gamma\rho}\epsilon_{\delta \lambda \rho \sigma}dB\wedge dx^{\sigma} \nonumber\\
    &+&\frac{1}{24 m_0 k}\star dA\wedge d\bigg[ (-g)^{1/2}\partial_{[\alpha}B_{\beta\gamma]}g^{\alpha(\mu}g^{\nu)\delta}g^{\beta\lambda}g^{\gamma\rho}\epsilon_{\delta \lambda \rho \sigma}dx^{\sigma} \bigg] \,.
\end{eqnarray}
From Eqs.~\eqref{definition_EMT} and \eqref{variation_metric}, the energy-momentum tensor in a generic form becomes:
\begin{eqnarray}\label{energy_momentum_tensor_generic}
    T^{\mu\nu}d^{4}x&=& \frac{1}{2}(-g)^{-1/2}g^{\mu\nu}\bigg[ dA\wedge \star dA+dB\wedge \star dB+\frac{1}{2m_0 k}\star dA\wedge d\star dB \bigg] \nonumber\\
    &+&\frac{(-g)^{-1/2}}{4m_0 k}\star dA\wedge d(g^{\mu\nu}\star dB) \nonumber\\
    &-&\frac{1}{2}\partial_{[\alpha}A_{\beta]}g^{\alpha(\mu}g^{\nu)\lambda}g^{\beta\rho}\epsilon_{\lambda \rho \gamma \delta}\bigg( dA\wedge dx^{\gamma}\wedge dx^{\delta}+\frac{1}{2m_0 k}dx^{\gamma}\wedge dx^{\delta}\wedge d\star dB \bigg) \nonumber\\
    &-&\frac{1}{6}\partial_{[\alpha]}B_{\beta\gamma}g^{\alpha(\mu}g^{\nu)\delta}g^{\beta\lambda}g^{\gamma\rho}\epsilon_{\delta \lambda \rho \sigma}dB\wedge dx^{\sigma} \nonumber\\
    &-&\frac{(-g)^{-1/2}}{12 m_0 k}\star dA\wedge d\bigg( (-g)^{1/2}\partial_{[\alpha}B_{\beta\gamma]}g^{\alpha(\mu}g^{\nu)\delta}g^{\beta\lambda}g^{\gamma\rho}\epsilon_{\delta \lambda \rho \sigma}dx^{\sigma} \bigg) \,.
\end{eqnarray}
On the other hand, we take the following properties of the fields $A$ and $B$:
\begin{equation}
    A_0=B_{\mu 0}=B_{0\mu}=0 \,.  
\end{equation}
%
%
Taking into account the components of the energy-momentum tensor in Eq.~\eqref{energy_momentum_tensor_generic} explicitly for the FLRW metric Eq~(\ref{RW1}), the expression is read in components as:
\begin{eqnarray}
    T^{00}&=& -\frac{1}{4a^2}\partial_0 A_i \partial_0 A_i-\frac{1}{18a^4}\partial_0 B_{ij}\partial_0 B_{ij}+\frac{1}{72m_0 k a^2}\partial_0 A_i \partial_0(a^5 \partial_0 B_{jk}\epsilon^{0ijk}), \\
    T^{0i}&=&0, \\
    T^{ij}&=&-\frac{1}{2a^5}\delta_{ij}\left[ \frac{a}{2}\partial_0 A_k \partial_0 A_k-\frac{1}{3a}\partial_0B_{kl}\partial_0 B_{kl}+\frac{a}{24m_0 k}\partial_0 A_k \partial_0 \left( a^3 \partial_0 B_{lm}\epsilon^{0lm}{}_{k} \right) \right] \nonumber\\
    &+&\frac{1}{48m_0 k}\frac{1}{a^2}\delta_{ij}\partial_{0}A_k \partial_0 \left( a^3 \partial_0 B_{lm}\epsilon^{lmk} \right)+\frac{1}{2a^4}\partial_0 A_k \delta_{k(i}\delta_{j)l}\left[ \partial_0 A_l+\frac{1}{12m_0 k}\partial_0 \left( a^3 \partial_0 B_{mn}\epsilon^{0mn}{}_{l} \right) \right] \nonumber\\
    &+& \frac{1}{4a^6}\partial_0B_{k(i|}\partial_0 B_{|j)k}+\frac{1}{36m_0 k}\frac{1}{a^2}\partial_0 A_k \partial_0 \left( \frac{1}{a^3}\partial_0 B_{l(i|}\epsilon_{0|j)kl} \right), \\
    T&:=& -\frac{1}{4a^5}\partial_0 A_i \partial_0 A_i+\frac{3}{4a^6}\partial_0 B_{ij}\partial_{0}B_{ij}-\frac{1}{48 m_0 k}\frac{1}{a^4}\partial_0 A_i \partial_0 \left( \frac{1}{a}\partial_0 B_{jk}\epsilon_{0ijk} \right) \nonumber\\
    &-&\frac{5}{144 m_0 k}\frac{1}{a^2}\partial_0 A_i \partial_0 \left( \frac{1}{a^3}\partial_0 B_{jk}\epsilon_{0ijk} \right) \,,
\end{eqnarray}
where $a(t)=\mathrm{exp}(H_0 t)$ is the explicit form of the scale factor and the last expression $T$ is the trace of the spatial components. The coupling of the 1-form $A$ and the 2-form $B$ in the effective action Eq. \eqref{faction} implies that $T^{ij}$ is a non-diagonal symmetric tensor. This cosmological model is analyzed by introducing the barotropic parameter $\omega$, that is,
\begin{equation}\label{barotropic}
\omega \equiv \frac{p}{\rho}=\frac{T}{3T^{00}} \,,
\end{equation}
where the choice of $3p=T$ and $\rho=T^{00}$ represents a perfect-fluid scheme. Here we recall that the fields $A$ and $B$ are given by the previous solutions Eq.\eqref{A_3} and Eq.\eqref{B_12}, respectively, and depend explicitly on the cosmological time $t$.  The complete system of solutions is a set of 4 equations; however, due to the analysis of the complete framework only 2 of them yield a relevant cosmological meaning, where the explicit expressions are presented in Appendix~\ref{pressure_density}: formulas $y_{1}$ (Eq.~(\ref{solutionA})) and $y_{2}$ (Eq.~(\ref{solutionB})). The role of $\omega$ is to characterize a dark energy framework in the following way: $-1<\omega<0$ where the quintessence scenario occurs. Here, the cosmic expansion is accelerating, but at a decreasing rate; then the phantom range occurs when $\omega<-1$, in this instance, the expansion of the universe is accelerating at an increasing rate. A combination of both occurrences yields the quintom realm, and this joint system allows us to cross the theoretical phantom divide line. Then, we present three different results where we have taken $H_{0}=67.70 \,\rm [km\, s^{-1}\,Mpc^{-1}]$~\cite{Planck:2018nkj}. Figures~\ref{fig:w_m_0} and ~\ref{fig:w_m_1} show the evolution of time $t$ and mass $m_{0}$ of $\omega (t)$ and $\omega (m_{0})$, respectively; where both plots are presented using $y_{1}$ (Eq.~(\ref{solutionA}). In contrast, figure~\ref{fig:w_m_10} only depicts $\omega (t)$ utilizing the solution $y_{2}$ (Eq.~(\ref{solutionB})). The first figure~\ref{fig:w_m_0} indicates a constant $\omega (t)\simeq -1$ since the selection of $m_{0}=0.727$ fixes such behavior to have a cosmological constant setting~\cite{Planck:2018nkj}. Then, figure~\ref{fig:w_m_1}, at $t=0$, exhibits a consistent growth, where $\omega (m_{0})$ can cross the phantom line (orange solid line) and for larger $m_{0}$ it tends to the quintessence scenario. Finally, in figure~\ref{fig:w_m_10}, with the mass scale parameter $m_{0}=0.027$, one observes that $\omega (t)$ oscillates from positive to negative values, it makes transit from quintessence to the phantom regimes; moreover, its maximum value reaches $\omega \simeq 1/3$ that corresponds to a radiation fluid.   

%
%
%
%
 



%
%
\section{Scale factor as a dynamical variable}\label{scale_factor}
In this section, we consider the scale factor as a dynamical variable in the 4-dimensional action \eqref{action_constrained}. Thus, the dynamical variables involved in this action are the scale factor $a$, and the fields $A_{\mu}$ and $B_{\mu\nu}$.  The variation of such action with respect to $a$ leads to the following:
\begin{eqnarray}\label{eq_for_a}
    &&\frac{1}{4m_0 k a^3}\bigg( \partial_0^2 A_3 \partial_0 B_{12}-\partial_0 A_3 \partial_0^2 B_{12}-\partial_0^2 A_2 \partial_0 B_{13}+\partial_0 A_2 \partial_0^2 B_{13}+\partial_0^2 A_1 \partial_0 B_{23}-\partial_0 A_1 \partial_0^2 B_{23} \bigg) \nonumber \\
    &-&\frac{1}{4}(\partial_0 A_1)^2-\frac{1}{4}(\partial_0 A_2)^2-\frac{1}{4}(\partial_0 A_3)^2-\frac{1}{2a^2}(\partial_0 B_{12})^2-\frac{1}{2a^2}(\partial_0 B_{13})^2-\frac{1}{2a^2}(\partial_0 B_{23})^2=0.
\end{eqnarray}
This equation can be regarded as an additional EOM associated with the scale factor $a$. Notice that this equation involves the three fields $A_{\mu}$, $B_{\mu\nu}$, and $a$. Taking into account the relationship among the components of fields $A$ and $B$ given in \eqref{C1}, \eqref{C2} and \eqref{C3}, the previous expression collapses to the following:
\begin{equation} \label{eoma}
    \frac{1}{4m_0 k}(-\Dot{A}_3 \Ddot{B}_{12}+\Ddot{A}_3 \Dot{B}_{12})= \frac{3}{4}a^3 (\Dot{A}_3)^2 +\frac{3}{2}a(\Dot{B}_{12})^2.
\end{equation}

\subsection{Solving the coupled differential equations system using series}
\begin{figure}[h]
\centering
\includegraphics[width=0.5\textwidth]{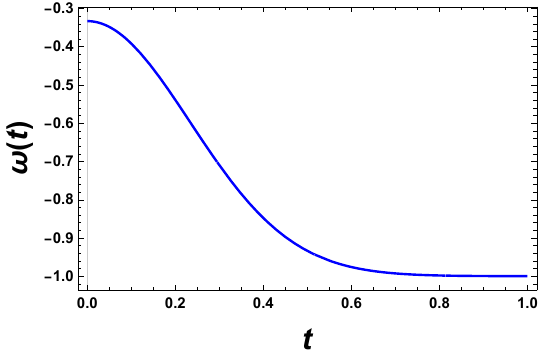}
\caption{Time evolution of the barotropic parameter $\omega (t)$ with $\mathcal{H}_{1}=A_{0}=B_{0}=H_{0}=67.70 \,\rm [km\, s^{-1}\,Mpc^{-1}]$~\cite{Planck:2018nkj}, $k_{A}=-10$, $k_{B}=0$, $a_{0}=1$, $k=1$, and the mass scale parameter $m_{0}=1.0$. Note that $\omega \rightarrow -1$ at larger $t$, which corresponds to a cosmological constant setting. Note that time is measured in reduced Planck units since $8\pi G=1$.}
\label{fig:w_m_2}
\end{figure}

\begin{figure}[h]
\includegraphics[width=0.49\textwidth]{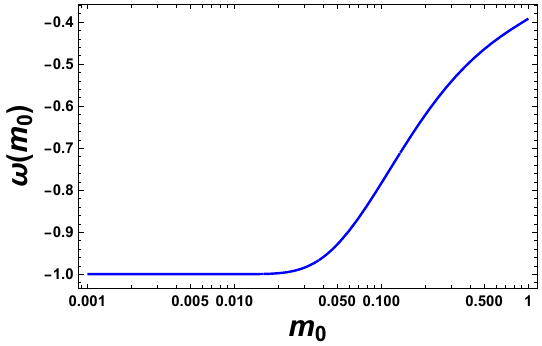}
\caption{In this figure we show the barotropic parameter $\omega (m_{0})$ in terms of the mass $m_{0}$ at $t\simeq 0$. Here we consider $\mathcal{H}_{1}=A_{0}=B_{0}=H_{0}=67.70 \,\rm [km\, s^{-1}\,Mpc^{-1}]$~\cite{Planck:2018nkj}, $k_{A}=-10$, $k_{B}=0$, $a_{0}=1$, $k=1$. Note that $\omega (m_{0})$ describes a cosmological constant setting when $m_{0}<0.02$, while for larger $m_{0}$ it tends to the quintessence scenario.}
\label{fig:w_m_3}
\end{figure}

In this section we proceed to solve the Eq.(\ref{geom}) as a first step, for both fields $A$ and $B$ since they have a similar shape, i.e., $E_{12}$ is $-a \partial_0 A_3$ or $\frac{1}{a} \partial_0 B_{12}$ which is the component $dx^1 \wedge dx^2$ of $\star dA$ and $dx^3$ of $\star d B$, respectively (we consider only one of the components of each form, since it is the same equation of motion for the other two nonzero components $A$ and similarly for $B$, as we saw previously in section \ref{Example}). Hence, explicitly the EOM have the form:
\begin{eqnarray}
    a\partial_0 \left( \frac{1}{a} \partial_0 \left( a \partial_0 A_3 \right)\right)+4m_0^2k^2 a \partial_0 A_3=0, \\
    -\frac{1}{a}\partial_0 \left( a \partial_0 \left( \frac{1}{a} \partial_0 B_{12} \right)\right)+4m_0^2k^2 \frac{1}{a} \partial_0 B_{12}=0.
\end{eqnarray}
Dividing these expressions by $a$ and $1/a$, respectively; we can factor out total derivatives which are equal to zero. This gives rise to the following set of second order differential equations:
%
\begin{eqnarray}
    \ddot{A}+\mathcal{H} \dot{A}+4 m_0^2 k^2 A=k_A, \label{eomA}\\
    \ddot{B}-\mathcal{H}\dot{B}-4 m_0^2 k^2 B=k_B, \label{eomB}
\end{eqnarray}
where we have dropped out the subindices such that $A_3=A$ and $B_{12}=B$, $k_A$ and $k_B$ are constants of integration and $\mathcal{H}:=\frac{\dot{a}}{a}$. Thus, the coupled differential equation system to solve in this section is given by \eqref{eoma}, \eqref{eomA} and \eqref{eomB}. To find a solution of this system\footnote{See Appendix \ref{Appendix_B} for the particular case when the fields $A$ and $B$ are weak.}, we will consider the following series expansion of the fields and their derivatives:
\begin{eqnarray}
    A=\sum_{k=0}^\infty A_k t^{k+r}, \\
    \dot{A}=\sum_{k=0}^\infty (k+r)A_k t^{k+r-1}, \\ 
    \ddot{A}=\sum_{k=0}^\infty (k+r)(k+r-1)A_k t^{k+r-2}.   
\end{eqnarray}
Since $\mathcal{H}$ is an unknown function, we also assume the power expansion given by:
\begin{equation}
    \mathcal{H}=\sum_{k=0}^\infty \mathcal{H}_k t^{k-1}.
\end{equation}
Substituting these expressions into the equation $(\ref{eomA})$ we obtain:
\begin{equation}
    \sum_{k=0}^\infty (k+r)(k+r-1)A_k t^{k+r-2}+\sum_{j,k=0}^\infty (k+r)A_k \mathcal{H}_j t^{j+k+r-2} +4m_0^2k^2\sum_{k=0}^\infty A_k t^{k+r}=k_A.
\end{equation}
Then the indicial equation ($k=0$) is as follows:
\begin{equation}
r(r-1) A_0+r\mathcal{H}_0A_0=0.
\end{equation}
Assuming $A_0 \neq0$, we have the following roots $r=0, 1-\mathcal{H}_0$. In this section, we will explore the solution associated with $r=0$. Then, the next lowest power after substituting $r=0$ gives $\mathcal{H}_0 A_1=0$. Considering the situation $A_1=0$, for the next orders we obtain:
\begin{eqnarray}\label{as}
    A_2=\frac{k_a-4m_0^2k^2 A_0}{2(1+\mathcal{H}_0)}, \\
    A_3=-\frac{2 \mathcal{H}_1 A_2}{3(2+\mathcal{H}_0)}, \\
    A_4= -\frac{3 \mathcal{H}_1 A_3+ 2 \mathcal{H}_2 A_2+ 4m_0^2k^2 A_2}{4(3+\mathcal{H}_0)}.\\
\end{eqnarray}
Similarly, for $B$, we consider the following anzats:
\begin{eqnarray}
    B=\sum_{k=0}^\infty B_k t^{k+r}, \\
    \dot{B}=\sum_{k=0}^\infty (k+r) B_k t^{k+r-1}, \\ 
    \ddot{B}=\sum_{k=0}^\infty (k+r)(k+r-1) B_k t^{k+r-2}.   
\end{eqnarray}
Thus, the equation of motion for $B$ is:
\begin{equation}
  \sum_{k=0}^\infty (k+r)(k+r-1)B_k t^{k+r-2} - \sum_{j,k=0}^\infty (k+r) B_k \mathcal{H}_j t^{j+k+r-2} - 4 m_0^2 k^2 \sum_{k=0}^\infty B_k t^{k+r}=k_B.  
\end{equation}
The indicial equation (i.e., the lowest power) for $B$ reads: 
\begin{equation}
    r(r-1)B_0-rB_0\mathcal{H_0} =0.
\end{equation}
Assuming $B_0 \neq 0$, the solutions are $r=0, 1+\mathcal{H_0}$, analogous to the equation of motion for $A$, we consider $r=0$ for the next lowest power which implies $\mathcal{H}_0 B_1=0$, then we consider $B_1=0$, therefore, solving the following powers we find the following values for coefficients:
\begin{eqnarray}\label{bs}
    B_2=\frac{k_B+4m_0^2k^2 B_0}{2(1-\mathcal{H}_0)}, \\
    B_3=\frac{2 \mathcal{H}_1 B_2}{3(2-R_0)}, \\
    B_4= \frac{3 \mathcal{H}_1 B_3+ 2 \mathcal{H}_2 B_2+ 4m_0^2k^2 B_2}{4(3-\mathcal{H}_0)}.\\
\end{eqnarray}
The coefficients ($\ref{as}$) and ($\ref{bs}$) are defined by those of $\mathcal{H}$; so we solve them using the motion equation for $a$ (\ref{eoma}):
\begin{equation}
    -\dot{A} \ddot{B} + \ddot{A} \dot{B}= 3 m_0 k a^3 (\dot{A})^2+6m_0 k a (\dot{B})^2.
\end{equation}
An important note is that the scale factor can be easily calculated, given that $\mathcal{H}=\frac{\dot{a}}{a}=\frac{d}{dt}\ln{a}$, so:
\begin{equation}
    a= \exp{\Big(\int \mathcal{H} dt\Big)}=\exp{\Big(\mathcal{H}_0 \ln{t} +\sum_{k=1}^\infty \frac{1}{k} \mathcal{H}_k t^k\Big)}.
\end{equation}
In order to avoid a singular solution at $t=0$, we set $\mathcal{H}_0=0$. Moreover, we need to expand $a$ in Taylor series to solve it in powers of $t$:
\begin{equation}
    a=\exp{\Big(\sum_{k=1}^\infty \frac{1}{k} \mathcal{H}_k t^k\Big)}=1+\sum_{k=1}^\infty \frac{1}{k} \mathcal{H}_k t^k+\frac{1}{2!}\left( \sum_{k=1}^\infty \frac{1}{k} \mathcal{H}_k t^k\right) \left( \sum_{j=1}^\infty \frac{1}{j} \mathcal{H}_j t^j\right)+ \cdots.
\end{equation}
We next substitute the power series for $A$, $B$, and $a$ into equation ($\ref{eoma}$). Considering $A_1=B_1=\mathcal{H}_0=0$, it follows that there are no relationships among the constants for orders zero and one, as both naturally equate to zero. At the $t^2$ order, we derive:
\begin{equation}
    -6 A_2 B_3 + 6 A_3 B_2= 12 m_0 k A_2^2+ 24 m_0 k B_2^2.
\end{equation}
This expression is not useful since it does not depend on $\mathcal{H}_i$, thus for order $t^3$ we obtain:
\begin{eqnarray}
    -16 A_2 B_4 +16 A_4 B_2-36 m_0 k A_2 A_3-72 m_0kB_2B_3=(36 m_0 k A_2^2+6 m_0 k B_2^2) \mathcal{H}_1.
\end{eqnarray}
Utilizing equations ($\ref{as}$) and ($\ref{bs}$), $\mathcal{H}'s$, $A's$, and $B's$ can be explicitly resolved through $A_0, B_0$, and $\mathcal{H}_1$, yielding:
\begin{eqnarray}
    A_2= \frac{k_A- 4 m_0^2 k^2 A_0}{2},\\
    A_3=-\frac{1}{3} \mathcal{H}_1 \left(\frac{k_A- 4 m_0^2 k^2 A_0}{2}\right),\\
    A_4=-\left(\frac{1}{12} \mathcal{H}_1^2+\frac{1}{6} \mathcal{H}_2+\frac{1}{3}m_0^2 k^2 \right)\left(\frac{k_A- 4 m_0^2 k^2 A_0}{2}\right),\\
    B_2=\frac{k_B+4m_0^2 k_2 B_0}{2},\\
    B_3=\frac{1}{3} \mathcal{H}_1\left(\frac{k_B+4m_0^2 k_2 B_0}{2} \right),\\
    B_4=\left(\frac{1}{12} \mathcal{H}_1^2+\frac{1}{6} \mathcal{H}_2 + \frac{1}{3}m_0^2 k^2\right) \left( \frac{k_B+4m_0^2 k_2 B_0}{2} \right),
\end{eqnarray}
where
\begin{equation}
    \mathcal{H}_2= -2 m_0^2 k^2-\frac{9}{2}m_0 k \mathcal{H}_1 \left[\frac{k_A+4 m_0^2 k^2 A_0}{k_B+ 4m_0^2 k^2 B_0} \right]-\frac{45}{8} m_0 k \mathcal{H}_1 \left[\frac{k_B+ 4 m_0^2 k^2 B_0}{k_A+4 m_0^2 k^2 A_0}\right].
\end{equation}
We proceed by explicitly listing the fields:
\begin{eqnarray}
A=A_0+\frac{k_A- 4 m_0^2 k^2 A_0}{2}t^2-\frac{1}{3}\mathcal{H}_1 \left(\frac{k_A- 4 m_0^2 k^2 A_0}{2}\right)t^3+ \left(\frac{1}{12} \mathcal{H}_1^2-\frac{1}{6} \mathcal{H}_2-\frac{1}{3}m_0^2 k^2 \right)\left(\frac{k_A- 4 m_0^2 k^2 A_0}{2}\right)t^4 \,, \label{nuevo_A}\\
    B=B_0+\frac{k_B+4m_0^2 k_2 B_0}{2} t^2+\frac{1}{3}\mathcal{H}_1\left(\frac{k_B+4m_0^2 k_2 B_0}{2} \right)t^3+\left(\frac{1}{12} \mathcal{H}_1^2+\frac{1}{6} \mathcal{H}_2 + \frac{1}{3}m_0^2 k^2\right) \left( \frac{k_B+4m_0^2 k_2 B_0}{2} \right)t^4 \,, \label{nuevo_B}
\end{eqnarray}
and the scale factor:
\begin{equation}\label{sol_a_complete}
    a(t)=\exp \left\{\mathcal{H}_1 t-\frac{1}{4}\left(4 m_0^2 k^2+\frac{9}{2}m_0 k \mathcal{H}_1 \left[\frac{k_A+4 m_0^2 k^2 A_0}{k_B+ 4m_0^2 k^2 B_0} \right]+\frac{45}{8} m_0 k \mathcal{H}_1 \left[\frac{k_B+ 4 m_0^2 k^2 B_0}{k_A+4 m_0^2 k^2 A_0}\right]\right)t^2 \right\}.
\end{equation}
In the expression for \eqref{sol_a_complete}, it is conventional, and without loss of generality, to equate $\mathcal{H}_1$ to the Hubble parameter $H_0$, similar to the de Sitter model. The higher order time terms may be perceived as corrections to the lowest order solutions for the fields $A$ and $B$, the mass $m_0$, and the constant $k$. These corrections rapidly diminish on the indicial conditions and the values of $m_0$ and $k$. With the barotropic parameter $\omega$ defined in \eqref{barotropic}, and using the previously obtained solutions for fields $A$ and $B$ from Eqs. \eqref{nuevo_A} and \eqref{nuevo_B}, we propose two different results, assuming $67.70 \,\rm [km\, s^{-1}\,Mpc^{-1}]$~\cite{Planck:2018nkj}. Figures~\ref{fig:w_m_2} and~\ref{fig:w_m_3} illustrate the evolution of $\omega(t)$ over time $t$ and $\omega(m_{0})$ with the mass parameter $m_{0}$, respectively; both are plotted using $\mathcal{H}_{1}=A_{0}=B_{0}=H_{0}=67.70 \,\rm [km\, s^{-1}\,Mpc^{-1}]$~\cite{Planck:2018nkj}, $k_{A}=-10$, $k_{B}=0$, $a_{0}=1$, and $k=1$. Figure~\ref{fig:w_m_2} shows that for smaller times $\omega$ lies in the quintessence regime, while for larger $t$ $\omega \simeq -1$, indicating a cosmological constant scenario. Then, figure~\ref{fig:w_m_3} reveals that at $t\simeq 0$, $\omega (m_{0})$ exhibits a consistent growth,transitioning from a cosmological constant ($\omega (m_{0})\simeq -1$) when $m_{0}<0.02$, to a quintessence framework for larger mass values of $m_{0}$. It is important to note that this behavior occurs primarily in the early stages.

\section{Conclusions}
\label{Conclusions}
%
%
In this work, we consider a $4$-dimensional effective $BF$ theory action given in~\cite{efremov2014universe} that arises after a Kaluza-Klein dimensional reduction of a $7$-dimensional $BF$ theory action, assuming the ansatz given by \eqref{ansatz}. We explicitly express the effective action and the EOM in coordinates. We study two different scenarios: the first consists of a toy model in which the FLRW metric is used, while in the second the analysis is performed with the scale factor $a(t)$ as a dynamical variable. The first case is examined in Sect. \ref{Example}, where the resulting cosmological model corresponds to a de Sitter universe, since the scale factor was fixed as $a(t)=\mathrm{exp}(H_0 t)$, with $H_{0}=67.70 \,\rm [km\, s^{-1}\,Mpc^{-1}]$ as the Hubble constant~\cite{Planck:2018nkj}. Here, the effective action may represent a quintom scenario. To analyze our solutions, we compute the barotropic parameter $\omega \equiv p/\rho$. Three different results are presented on two distinct relevant solutions: $y_{1}$ (Eq.~(\ref{solutionA})) and $y_{2}$ (Eq.~(\ref{solutionB})). Using $y_{1}$, the figures~\ref{fig:w_m_2} and ~\ref{fig:w_m_3} show the evolution of time $t$ and the mass evolution $m_{0}$ of $\omega (t)$ and $\omega (m_{0})$, respectively. We fix  $m_{0}=0.727$ to stablish the cosmological constant scenario ~\cite{Planck:2018nkj}; meanwhile, at $t=0$, $\omega (m_{0})$ can cross the phantom line (orange solid line) and for larger $m_{0}$ it tends to the quintessence scenario. In contrast, the figure~\ref{fig:w_m_10} only depicts $\omega (t)$ utilizing the solution $y_{2}$ (Eq.~(\ref{solutionB}) and $m_{0}=0.027$. This case is more relevant since $\omega (t)$ oscillates from positive to negative values, it travels from quintessence to phantom regimes; remarkably, its maximum value reaches $\omega \simeq 1/3$, which corresponds to a radiation fluid. Thus, this result yields the most exotic matter content. Extraordinarily, this toy model can travel across various cosmic expansion epochs. Subsequently, in Sect. \ref{scale_factor} the second case is considered. Here, the scalar factor is regarded as a dynamical variable. At this stage, we solved a coupled differential equation system using a series methodology for fields $A$, $B$, and the scale factor $a$. Then, the barotropic parameter $\omega$ is computed to analyze the solutions. Two different results are presented from the relevant solutions of $A$ and $B$, as given by Eqs.~\eqref{nuevo_A} and~\eqref{nuevo_B}. Figures~\ref{fig:w_m_0} and~\ref{fig:w_m_1} show the evolution of time $t$ and the evolution of mass $m_{0}$ of $\omega (t)$ and $\omega (m_{0})$, respectively. In both cases $\omega$ can transit from a cosmological constant ($\omega \simeq -1$) to a quintessence framework ($\omega > -1$). Hence, this model cannot cross the theoretical phantom divide line. However, these findings might indicate the presence of exotic matter.

Finally, we expect that this novel cosmology from the $BF$ theory captures the attention from the theoretical framework standpoint. We consider extending our work by including the CMPR action~\cite{Capovilla_2001} in order to perform the canonical analysis coupling with gravity in a $BF$ framework.
  
\section{Acknowledgments}
\label{Acknowledgments}
This work was supported by the CONAHCYT Network Project No. 376127 {\it Sombras, lentes y ondas gravitatorias generadas por objetos compactos astrofísicos}. R.S.S. would like to thank the Isaac Newton Institute for Mathematical Sciences for support and hospitality during the program ``Balck holes: bridges between number theory and holgraphic quantum information'', when part of the work on this paper was undertaken. Also R.S.S. would like to thank N. Cabo-Bizet and A. Martínez-Merino for their fruitful comments. R.E. and R.H.J. are supported by CONAHCYT Estancias Posdoctorales por M\'{e}xico, Modalidad 1: Estancia Posdoctoral Acad\'{e}mica and by SNI-CONAHCYT. C.M. and R.S.S want to thank SNI-CONAHCYT, PROINPEP-UDG, PROSNI-UDG and SEP-PRODEP.
\appendix

\section{Differential forms}\label{Appendix_A}

In this appendix, we write some useful definitions of differential forms that were used to write down the action and the EOM in their most general form. We follow the references very closely~\cite{Nakahara:2003nw, Frankel:2011}. A $p$-form $\alpha$ over the $n$-dimensional manifold $M$ in coordinates is written as:
\begin{equation}\label{p_form}
    \alpha= \frac{1}{p!}\alpha_{\mu_1 \ldots \mu_p} dx^{\mu_1} \wedge \cdots \wedge dx^{\mu_p} \,,
\end{equation}
where the coefficients $\alpha_{\mu_1 \ldots \mu_p}$ are symmetric in skewness (it can be regarded a tensor that is totally symmetric in skewness) and the indexes $\mu_i$ are arranged over $1,2 \ldots, n$ and are functions of the manifold $M$. The term $dx^{\mu_1} \wedge \cdots \wedge dx^{\mu_p}$ is the base expansion of the $p$-forms in terms of the coordinate $x^\mu$'s, here $\wedge$ is named the wedge product and represents the operation among the $1$-forms to form a skew-symmetric object. Finally, the space of $p$ forms spanned by the basis $dx^{\mu_1} \wedge \cdots \wedge dx^{\mu_p}$ is denoted by $\Omega^p (M)$.

The exterior derivative is defined as a map $d: \Omega^p(M) \rightarrow \Omega^{p+1}(M)$  explicitly given in coordinates as:
\begin{equation}
d \alpha=\frac{1}{p!} (\partial_\nu \, \alpha_{\mu_1 \ldots \mu_p}) dx^\nu \wedge dx^{\mu_1} \wedge \cdots \wedge dx^{\mu_p} \,, 
\end{equation}
where $\alpha$ is the $p$-form denoted previously in \eqref{p_form}. On the other hand, the Hodge dual map 'star' $\star: \Omega^p \rightarrow \Omega^{n-p}$, is defined by:
\begin{eqnarray}\label{hodge}
    \star \alpha &:=& \star\left(\frac{1}{p!}\alpha_{\mu_{1}\cdots \mu_{p}}dx^{\mu_1}\wedge \cdots \wedge dx^{\mu_p} \right) \nonumber \\
    &=& \frac{\sqrt{-g}}{p!(n-p)!}\alpha_{\mu_{1}\cdots \mu_p}\epsilon^{\mu_1 \cdots \mu_p}{}_{\nu_1\cdots \nu_{n-p}}dx^{\nu_1}\wedge \cdots \wedge dx^{\nu_{n-p}} \,,
\end{eqnarray}
where $\epsilon^{\mu_1\cdots \mu_p}{}_{\nu_1 \cdots \nu_{n-p}}=g^{\mu_1\rho_1}\cdots g^{\mu_p\rho_p}\epsilon_{\rho_1 \cdots \rho_p \nu_1 \cdots \nu_{n-p}}$, here $n$ is the dimension. For the Lorentzian case, the following property holds for the Hodge star operator:
\begin{eqnarray}
    \star^2 \alpha &=& (-1)^{1+p(n-p)}\alpha \,.
\end{eqnarray}
In order to compute the expressions easily, we define an inner product in the differential space form defined as:
\begin{equation}
(\alpha,\beta):=\int_{M_4}\alpha \wedge \star \beta \,,
\end{equation}\label{effective_action}
with $\alpha, \beta \in\Omega^{p}(M)$. Due to the linearity of $d$ and the inner product, we define the adjoint operator of $d^\dagger$ given by the expression:
\begin{equation}
      (d\alpha,\beta)=(\alpha,d^{\dagger}\beta) \,,
\end{equation}
and explicitly $d^\dagger$ is given by
\begin{equation}\label{dag}
    d^{\dagger} = (-1)^{n(p+1)}\star d\star \,.
\end{equation}

\section{Solution for weak fields $A$ and $B$} \label{Appendix_B}
The differential equation system, initially represented by $\eqref{eoma}$, $\eqref{eomA}$, and $\eqref{eomB}$, simplifies under the assumption of weak fields $A$ and $B$ by neglecting quadratic field terms. Thus, the system to be solved reads
\begin{eqnarray}
\label{Eq_1_frob}-\Dot{A}\Ddot{B}+\Ddot{A}\Dot{B}&=&0, \\
\label{Eq_2_frob}a\Ddot{A}+\Dot{a}\Dot{A}+4m_0k^2 aA&=&ak_A, \\
\label{Eq_3_frob}a\Ddot{B}-\Dot{a}\Dot{B}-4m_0k^2aB&=&ak_B.
\end{eqnarray}
Notably, the scale factor $a$ is absent in \eqref{Eq_1_frob} once the weak field approximation is applied. To find a solution of this system, we use a series method which is a Frobenius-like approach (see, for instance, \cite{Arfbook}). We assume that the fields $A$, $B$, and $a$ take the following form:
\begin{eqnarray}
    \label{A}A&=& \sum_{\lambda=0}^{\infty}g_{\lambda}t^{r+\lambda}, \quad g_{0}\neq 0, \\
    \label{B}B&=& \sum_{m=0}^{\infty}b_{m}t^{r+m}, \quad b_0\neq 0, \\
    \label{scalar_factor_a}a&=&\sum_{n=0}^{\infty}c_{n}t^{r+n}, \quad c_0\neq 0,
\end{eqnarray}
where the exponent $r$ and the coefficients $g_{\lambda}$, $b_{\lambda}$, and $c_{\lambda}$ are still undetermided. By differentiating twice, we obtain:
\begin{eqnarray}
 \Dot{A}&=& \sum_{\lambda=0}^{\infty}g_{\lambda}(r+\lambda)t^{r+\lambda-1}, \\
 \Ddot{A}&=& \sum_{\lambda=0}^{\infty}g_{\lambda}(r+\lambda)(r+\lambda-1)t^{r+\lambda-2},\\
 \Dot{B}&=& \sum_{m=0}^{\infty}b_{m}(r+m)t^{r+m-1}, \\
 \Ddot{B}&=& \sum_{m=0}^{\infty}b_{m}(r+m)(r+m-1)t^{r+m-2}, \\
 \Dot{a}&=& \sum_{n=0}^{\infty}c_{n}(r+n)t^{r+n-1}, \\
 \Ddot{a}&=& \sum_{n=0}^{\infty}c_{n}(r+n)(r+n-1)t^{r+n-2}.
\end{eqnarray}
Substituting the previous series into \eqref{Eq_1_frob}, \eqref{Eq_2_frob}, and \eqref{Eq_3_frob} yields:
\begin{eqnarray}
    0&=&0, \\
    \sum_{n=0}^{\infty}\sum_{\lambda=0}^{\infty}c_n g_{\lambda}\left[ (2r+\lambda+n-1)(r+\lambda)t^{2r+\lambda+n-2}+4m_0^2 k^2 t^{2r+\lambda+n} \right]&=&0, \\
    \sum_{m=0}^{\infty}\sum_{n=0}^{\infty}b_m c_n\left[ (m-n-1)(r+m)t^{2r+m+n-2}-4m_0^2 k^2 t^{2r+m+n} \right]&=&0.
\end{eqnarray}
Notice that \eqref{Eq_1_frob} is trivially fulfilled after the substitution of \eqref{A} and \eqref{B}. This is a consequence of the form of the fields $A$ and $B$ given in \eqref{A} and \eqref{B} respectively, and the absence of the scale factor $a$ in \eqref{Eq_1_frob}. Thus, at this stage, we can interpret the above system as two differential equations for three unknowns. In order to find a dynamical solution, we must fix one field to solve the system. Since the fields $A$ and $B$ are weak, without loss of generality two different cases are proposed for the scalar function without loss of generality. We highlight that the freedom to fix this function comes from the assumption that the fields $A$ and $B$ are weak and the form of them given in \eqref{A} and \eqref{B}, respectively.

%
\begin{itemize}

\item \textit{First case for the scalar function}.
The form we consider here for the scalar function reads as follows:
\end{itemize} \begin{equation}\label{scalar_factor_exp}
     a(t)=e^{H_0 t},
 \end{equation}
 with $H_0$ as the Hubble constant parameter. Substituting into \eqref{Eq_1_frob}, \eqref{Eq_2_frob}, and \eqref{Eq_3_frob}, taking into account the homogeneous system and considering only the lowest power of $t$ appearing in each equation, we arrive to the following set of indicial equations
 \begin{eqnarray}
     0&=&0, \\
     r(2r-1)&=&0, \\
     r&=&0.
 \end{eqnarray}
Thus, as the three previous equations are coupled among each other, the solution of such a system leads to:
\begin{equation}\label{indicial_equation}
    r=0.
\end{equation}
Therefore, \eqref{indicial_equation} is the indicial equation for the coupled system. Comparing the scale factor \eqref{scalar_factor_exp} with \eqref{scalar_factor_a}, we find that the coefficients are given by:
\begin{eqnarray}
    c_n=\frac{H_0^n}{n!}.
\end{eqnarray}
Demanding that the remaining coefficients in \eqref{Eq_2_frob} and \eqref{Eq_3_frob} vanish, leads to the following recurrence relations for $g_{j}$ and $b_j$, respectively as follows:
\begin{eqnarray}
g_{j+1}&=& -g_j \frac{1}{H_0}\frac{4m_0^2 k^2}{(2j+1)}, \\
b_{j+1}&=& -b_j \frac{1}{H_0}4m_0^2 k^2.
\end{eqnarray}
Thus, the complete form of the weak fields $A$ and $B$ (taking into account the particular solution of each of them) gives rise to the following result:
\begin{eqnarray}
    A(t)&=& \frac{k_A}{4m_0^2 k^2}+ g_0 \bigg[ 1-\frac{4m_0^2 k^2}{H_0}t+\frac{(4m_0^2 k^2)^2}{3H_0^2}t^2- \cdots \bigg], \\
    B(t)&=& -\frac{k_B}{4m_0^2 k^2}+ b_0 \bigg[ 1-\frac{4m_0^2 k^2}{H_0}t+\frac{(4m_0^2 k^2)}{H_0^2}t^2- \cdots \bigg].
\end{eqnarray}

\begin{itemize}

\item \textit{Second case for the scalar function.}
Analyzing the case when the scale factor is $a(t)\sim \mathrm{sinh}^{2/3}(H_0 t)$ is interesting because in series, for the radiation-dominated era~\cite{Scale_factor_with_Dark_Energy}, it reduces to $a(t)\sim (H_0 t)^{2/3}$. However, in our approach, the computations are not feasible to carry out. Hence, we consider this particular form for the scalar function:
\end{itemize}
\begin{equation}\label{form_2_scalar_function}
    a(t)=\mathrm{sinh}(H_{0}t).
\end{equation}
In order to avoid singularities in the recurrence relations of fields $A$ and $B$, our approach will differ slightly from the previous case. We begin by assuming that the fields $A$, $B$, and $a$ have the following form:
\begin{eqnarray}
    A&=&\sum_{\lambda=0}^{\infty} g_{\lambda}t^{r+\lambda} \quad g_0\neq 0, \\
    B&=& \sum_{m=0}^{\infty} b_{2m}t^{r+2m} \quad b_{0}\neq 0,\\
    \label{a_sinh_serie}a&=& \sum_{n=0}^{\infty}c_{2m+1}t^{r+2m+1} \quad c_{1}\neq 0.
\end{eqnarray}
It is remarkable that the series for the fields $B$ and $a$ are even and odd, respectively\footnote{Attempting an odd series for $B$ implies that solving the set of indicial equations given by \eqref{Eq_1_frob}, \eqref{Eq_2_frob}, and \eqref{Eq_3_frob} is not possible to solve.}. The even series for $B$ is required to avoid singularities in the recurrence relation of its coefficients, whereas the odd series for $a$ is due to the fact that the scalar function is an odd function given by \eqref{form_2_scalar_function}. Substituting the previous expressions for $A$, $B$ and $a$ into \eqref{Eq_2_frob} and \eqref{Eq_3_frob}, and considering the lowest order of $t$, the set of indicial equations collapses to:
\begin{eqnarray}
    r^2&=&0, \\
    r&=&0,
\end{eqnarray}
which solution is $r=0$. Demanding that the remaining coefficients in \eqref{Eq_2_frob} and \eqref{Eq_3_frob} vanish, we obtain the recurrence relations for the coefficients for $A$ and $B$ as follows:
\begin{eqnarray}
    \label{rec_a} g_{j+2}&=&-g_j \frac{4m_0^2 k^2}{(j+2)(3j+2)} \quad \mathrm{for}\hspace{1mm} j\hspace{1mm} \mathrm{even}, \\
    \label{rec_b} b_{2(j+1)}&=&-b_{2j}\frac{4m_0^2 k^2}{(j+1)(3j+2)}.
\end{eqnarray}
Moreover, bearing in mind that $r=0$, and comparing \eqref{form_2_scalar_function} with \eqref{a_sinh_serie}, leads to the following:
\begin{equation}
c_{2j+1}=\frac{(H_0)^{2j+1}}{(2j+1)!}.
\end{equation}
It is worth noting that the coefficients in \eqref{a_sinh_serie} are independent of prior recurrence formulas, implying that the solutions for $A$ and $B$ exclude the Hubble parameter $H_0$. Furthermore, the odd terms in \eqref{rec_a} and \eqref{rec_b} are zero. Thus, the resulting expressions for fields $A$ and $B$ are given as follows:
\begin{eqnarray}
    A(t)&=&\frac{k_A}{4m_0^2 k^2}+g_0\left[  1-\frac{4m_0^2 k^2}{4}t^2+\frac{(4m_0^2 k^2)^2}{128}t^4-\cdots \right], \\
    B(t)&=&-\frac{k_B}{4m_0^2 k^2}+b_0\left[ 1-\frac{4m_0^2 k^2}{2}t^2+\frac{(4m_0^2 k^2)^2}{20}t^4-\cdots \right].
\end{eqnarray}

\section{Pressure and density}\label{pressure_density}

\begin{figure}[h]
\centering
\includegraphics[width=0.5\textwidth]{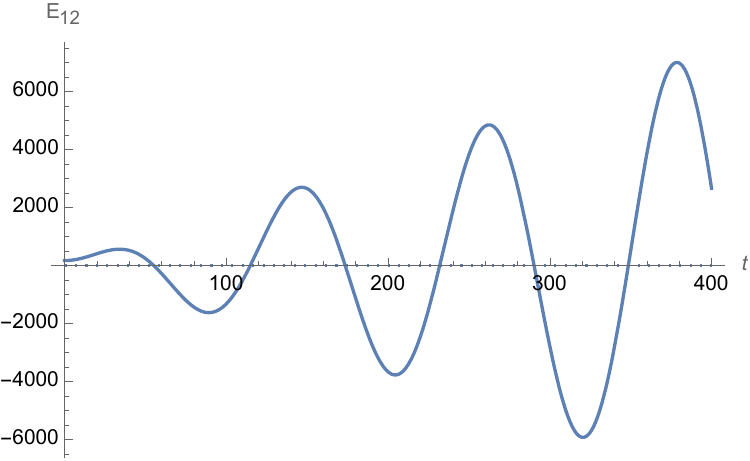}
\caption{Solution $y_{2}$ (Eq.~(\ref{solutionB}), which is a linear combination of the solutions Eq.\eqref{Sol_3} and Eq.\eqref{Sol_4 }, with the mass scale parameter $m_0=0.727$, $k=1$, and Hubble parameter $H_{0}=67.70 \,\rm [km\, s^{-1}\,Mpc^{-1}]$.}
\label{fig:y_2}
\end{figure}

In this appendix, we write down explicitly for the FLRW model the expression used to compute the barotropic parameter $\omega$ (Eq.~\eqref{barotropic}). First, for the solution $y_{1}$, we consider the solution in Eq.\eqref{Sol_1}
\begin{equation}\label{solutionA}
y_{1}(t)=c_1 e^{\frac{1}{2} t (H_0-f)} \,, 
\end{equation}
where $c_1$ is a constant and we have defined $f=\sqrt{H_0^2-16 k^2 m_0^2}$. Hence, the pressure reads:
\begin{eqnarray} \label{pressure1}
 p_{1} &=& \frac{\left[e^{H_0 t} \left(253 H_0 \left(H_0-f \right)-1880 k^2 m_0^2\right)-288 k^2 m_0^2\right] e^{-t \left(f+6 H_0\right)}}{384 k^2 m_0^2} \,,
\end{eqnarray}
while the density is given by:
\begin{eqnarray}\label{density1}
\rho_{1}=& & \frac{\left[8 k^2 m_0^2 \left(72 k m_0 e^{2 H_0 t}-7\right)+9 H_0 \left(f-H_0\right)\right] e^{-t \left(f+3 H_0\right)}}{64 k^2 m_0^2} \,.
\end{eqnarray}
%
%
Next, for the solution $y_{2}$, we consider a linear combination of the solutions Eq.\eqref{Sol_3} and Eq.\eqref{Sol_4}:
\begin{equation}\label{solutionB}
y_{2}(t) =-\frac{c_2}{f^2} (f t+1) e^{\frac{1}{2} t (H_0-f)}+\frac{c_3}{f^2} (f t-1) e^{\frac{1}{2} t (f+H_0)} \,,   
\end{equation}
with $c_2$ and $c_3$ as constants the pressure becomes: 
\begin{eqnarray} \label{pressure2}
 & &p_{2} = \frac{-1}{589824 k^8 m_0^8 f^4} \Bigg\{ 27 \left(f+H_0\right)^4 e^{-t \left(f+6 H_0\right)} \Big[-H_0^3 t+H_0^2 \left(t f+1\right)-H_0 \left(f-16 k^2 m_0^2 t\right)  \nonumber \\ 
 & & -8 k^2 m_0^2 \left(t f+1\right)+e^{t f} \left(H_0^3 t+H_0^2 \left(1-t f\right)-H_0 \left(f+16 k^2 m_0^2 t\right)+8 k^2 m_0^2 \left(t f-1\right)\right) \Big]^2 \nonumber \\
 & &-5184 k^2 m_0^2 e^{-t \left(f+5 H_0\right)} \Big[H_0^3-2 H_0 k^2 m_0^2 \left(3 t f+11\right)-H_0^2 \left(f-6 k^2 m_0^2 t\right)+10 k^2 m_0^2 f \nonumber \\ 
 & &  +e^{t f} \left(H_0^3+H_0^2 \left(f+6 k^2 m_0^2 t\right)+2 H k^2 m_0^2 \left(3 t f-11\right)-2 k^2 m_0^2 \left(5 f+48 k^2 m_0^2 t\right)\right)-96 k^4 m_0^4 t\Big]^2 \nonumber \\
 & &+24 k^2 m_0^2 e^{-t \left(f+5 H_0\right)}\left(f+H_0\right)^2 \Big[ H_0^3 t-H_0^2 \left(t f+1\right)+H_0 \left(f-16 k^2 m_0^2 t\right)+8 k^2 m_0^2 \left(t f+1\right)  \nonumber \\ 
 & & +e^{t f} \left(-H_0^3 t+H_0^2 \left(t f-1\right)+H_0 \left(f+16 k^2 m_0^2 t\right)-8 k^2 m_0^2 \left(t f-1\right)\right) \Big] \Big[4 k^2 m_0^2 \left(3 t f-1\right)  \nonumber \\ 
 & &  +e^{t f} \left(H_0^2-4 k^2 m_0^2 \left(3 t f+1\right)\right)+H_0^2 \Big] \nonumber \\
 & & +20 \left(f+H_0\right)^2 e^{-t \left(f+5 H_0\right)} \Big[ -H_0^3 t+H_0^2 \left(t f+1\right)-H_0 \left(f-16 k^2 m_0^2 t\right)-8 k^2 m_0^2 \left(t f+1\right)  \nonumber \\ 
 & & +e^{t f} \left(H_0^3 t+H_0^2 \left(1-t f\right)-H_0 \left(f+16 k^2 m_0^2 t\right)+8 k^2 m_0^2 \left(t f-1\right)\right) \Big] \Big[H_0^4-6 H^2_0 k^2 m_0^2 \left(t f+4\right)  \nonumber \\ 
 & &   +2 H_0 \left(5 k^2 m_0^2 f-48 k^4 m_0^4 t\right)+8 k^4 m_0^4 \left(1-3 t f\right)-H_0^3 \left(f-6 k^2 m_0^2 t\right)+e^{t f} \left(H_0^4+6 H_0^2 k^2 m_0^2 \left(t f-4\right) \right. \nonumber \\ 
 & &  \left. -2 H_0 \left(5 k^2 m_0^2 f+48 k^4 m_0^4 t\right)+8 k^4 m_0^4 \left(3 t f+1\right)+H_0^3 \left(f+6 k^2 m_0^2 t\right)\right)\Big] \Bigg\} \,,
\end{eqnarray}
and the density is:
\begin{eqnarray}\label{density2}
& &\rho_{2}= -\frac{1}{196608 k^8 m_0^8 f^4} \Bigg\{ 27 \left(f+H_0\right)^4 e^{-t \left(f+3 H_0\right)} \Big[-H_0^3 t+H_0^2 \left(t f+1\right)-H_0 \left(f-16 k^2 m_0^2 t\right)-8 k^2 m_0^2 \left(t f+1\right)  \nonumber \\
& & +e^{t f} \left(H_0^3 t+H_0^2 \left(1-t f\right)-H_0 \left(f+16 k^2 m_0^2 t\right)+8 k^2 m_0^2 \left(t f-1\right)\right)\Big]^2 \nonumber \\ 
& & +384 k^2 m_0^2 e^{-t \left(f+3 H_0\right)} \Big[H_0^3-2 H k^2 m_0^2 \left(3 t f+11\right)-H_0^2 \left(f-6 k^2 m_0^2 t\right)+10 k^2 m_0^2 f  \nonumber \\ 
& &  +e^{t f} \left(H_0^3+H_0^2 \left(f+6 k^2 m_0^2 t\right)+2 H_0 k^2 m_0^2 \left(3 t f-11\right)-2 k^2 m_0^2 \left(5 +48 k^2 m_0^2 t\right)\right)-96 k^4 m_0^4 t\Big]^2  \nonumber \\
& & +32 k^2 m_0^2 \left(f+H_0\right)^2 e^{-t \left(f+3 H_0\right)} \Big[ H_0^3 t-H_0^2 \left(t f+1\right)+H_0 \left(f-16 k^2 m_0^2 t\right)+8 k^2 m_0^2 \left(t f+1\right) \nonumber \\ 
& &  +e^{t f} \left(-H_0^3 t+H_0^2 \left(t f-1\right)+H_0 \left(f+16 k^2 m_0^2 t\right)-8 k^2 m_0^2 \left(t f-1\right)\right) \Big] \cdot \nonumber \\ 
& & \cdot \Big[4 k^2 m_0^2 \left(3 t f-1\right)+e^{t f} \left(H_0^2-4 k^2 m_0^2 \left(3 t f+1\right)\right)+H_0^2\Big] \nonumber \\
& & +1152 k^3 m_0^3 \left(f+H_0\right)^2 e^{-t \left(f+H_0\right)} \Big[H_0^3 t-H_0^2 \left(t f+1\right)+H_0 \left(f-16 k^2 m_0^2 t\right)+8 k^2 m_0^2 \left(t f+1\right) \nonumber \\ 
& & +e^{t f} \left(-H_0^3 t+H_0^2 \left(t f-1\right)+H_0 \left(f+16 k^2 m_0^2 t\right)-8 k^2 m_0^2 \left(t f-1\right)\right)\Big] \cdot \nonumber \\
& & \cdot \Big[ 4 k^2 m_0^2 \left(3 t f-1\right)+e^{t f} \left(H_0^2-4 k^2 m_0^2 \left(3 t f+1\right)\right)+ H_0^2 \Big]\Bigg\} \,.
\end{eqnarray}
\begin{figure}[h]
\centering
\includegraphics[width=0.5\textwidth]{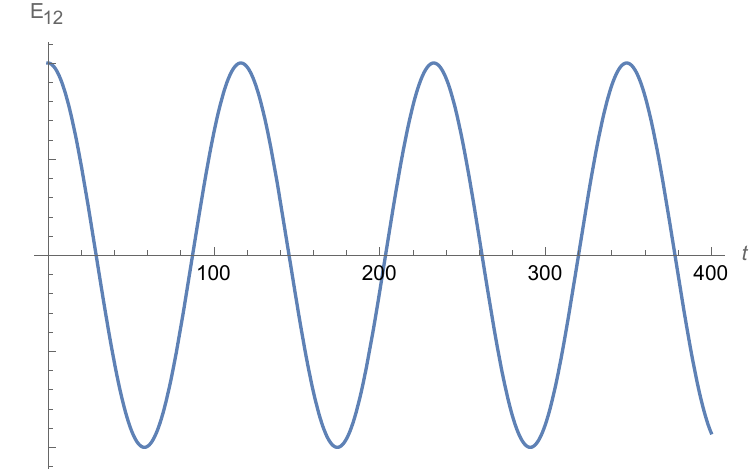}
\caption{Solution $y_{1}$ (Eq.~(\ref{solutionA}), with the mass scale parameter $m_0=0.727$, $k=1$, and Hubble parameter $H_{0}=67.70 \,\rm [km\, s^{-1}\,Mpc^{-1}]$.}
\label{fig:y_1}
\end{figure}
In figure \ref{fig:y_1}, we observe that the amplitude of the solution $y_1(t)$ is constant over time, while for the second solution $y_2$, its respective graph is in figure \ref{fig:y_2} and its amplitude is increasing over time.

\section{Conserved quantities}
\label{Conserved quantities}

In order for completeness of our work, we write down two additional equations associated with conserved quantities and the topological structure of the space-time. Consider Eq.~\eqref{EOM_1}, this expression leads to the following formula:
\begin{eqnarray}
    m_0 k \star dB-\star d\star dA-\frac{1}{4m_0 k}\star d \star d \star dB=0 \,,
\end{eqnarray}
then factoring out $\star d$, we have:
\begin{equation}\label{eq_1}
    \star d\left( m_0 k B-\star dA-\frac{1}{4m_0 k}d^{\dagger}dB \right)=0 \,,
\end{equation}
where we explicitly use the definition ($\ref{dag}$) considering $p=3$ and $n=4$. This implies that $m_0 k B-\star dA-\frac{1}{4m_0 k}d^{\dagger}dB$ is a closed $2$-form, which possible solution~{\footnote{In a more general way this could be a non trivial element of the second cohomology group of $M_4$.}} could be a constant form or in a more general way an exact form, i.e.:
\begin{equation}
    dC=m_0 k B-\star dA-\frac{1}{4m_0 k}d^{\dagger}dB \,,
\end{equation}
that trivially satisfies Eq.~(\ref{eq_1}) where $C\in \Omega^{1}(M)$, since $d^2$ over any form is zero. In a similar way to the case before, Eq.~\eqref{EOM_2} reduces to:
\begin{equation}
-m_0 k \star dA-\star d \star dB+\frac{1}{4m_0 k}\star d \star d \star dA=0 \,,
\end{equation}
implying that:
\begin{equation}
\star d \left( -m_0 k A-\star dB+\frac{1}{4m_0 k}d^{\dagger}dA \right)=0 \,.
\end{equation}
Analogously as in the previous case, we use $(\ref{dag})$ (setting $p=2$ and $n=4$) whose possible solution is given by:
\begin{equation}
dD=-m_0 k A-\star dB+\frac{1}{4m_0 k}d^{\dagger}dA \,,
\end{equation}
with $D\in \Omega^{0}(M)$ (that is, $D$ is a scalar function on $M$). But in general this expression is not necessarily cohomologically trivial, i. e. the equations could belong in general to a non-trivial cohomology group, thus this expression is sensible to the topology of the space-time.

\bibliography{bf_action}

\end{document}